\documentclass[fleqn,10pt]{wlscirep}
\usepackage{relsize}
\title{PopRank: Ranking pages' impact and users' engagement on Facebook}

\author[1,*]{Andrea Zaccaria}
\author[2]{Michela del Vicario}
\author[3,1]{Walter Quattrociocchi}
\author[1,4]{Antonio Scala}
\author[5,1]{Luciano Pietronero}

\affil[1]{Istituto dei Sistemi Complessi (ISC)-CNR, UOS Sapienza, Roma, Italy}
\affil[2]{IMT School for Advanced Studies Lucca, Italy}
\affil[3]{Ca' Foscari University of Venice, Italy}
\affil[4]{LIMS London Institute of Mathematical Sciences, London, UK}
\affil[5]{Dipartimento di Fisica, Sapienza Universit\`a di Roma, Italy}

\affil[*]{and.zaccaria@gmail.com}

\begin{abstract}
Users online tend to acquire information adhering to their system of beliefs and to ignore dissenting information. Such dynamics might affect page popularity. In this paper we introduce an algorithm, that we call PopRank, to assess both the \textit{Impact} of Facebook pages as well as users’ \textit{Engagement} on the basis of their mutual interactions. The ideas behind the PopRank are that i) high \textit{impact} pages attract many users with a low \textit{engagement}, which means that they receive comments from users that rarely comment, and ii) high \textit{engagement} users interact with high \textit{impact} pages, that is they mostly comment pages with a high popularity. The resulting ranking of pages can predict the number of comments a page will receive and the number of its posts. Pages’ \textit{impact} turns out to be slightly dependent on pages’ informative content (e.g., science vs conspiracy) but independent of users’ polarization. 
\end{abstract}
\begin{document}

\flushbottom
\maketitle
%
%
\thispagestyle{empty}

\section*{Introduction}

As of the third quarter of 2017, Facebook (FB) had 2.07 billion monthly active users~\cite{statista2018}, leading the rank of most popular social networking sites in the world. In the meantime, Oxford Dictionaries announced ``post-truth'' as the 2016 international Word of the Year~\cite{oxford}. Defined as an adjective ``relating to or denoting circumstances in which objective facts are less influential in shaping public opinion than appeals  to emotion and personal belief'', the term has been largely used in the context of the Brexit and Donald Trump's election in the United States and benefited from the rise of social media as news source. Indeed, Internet changed the process of knowledge production in an unexpected way. The advent of social media and microblogging platforms has revolutionized the way users access content, communicate and get informed. People can access to an unprecedented amount of information --only on FB more than 3M posts are generated per minute~\cite{allen2017what}-- without the intermediation of journalists or experts, thus actively participating in the diffusion as well as the production of content. Social media have rapidly become the main information source for many of their users: over half (51\%) of US users now get news via social media~\cite{newman2017reuters}. However, recent studies found that confirmation bias --i.e., the human tendency to acquire information adhering to one's system of beliefs-- plays a pivotal role in information cascades~\cite{del2016spreading}. Selective exposure has a crucial role in content diffusion and facilitates the formation of echo chambers --groups of like-minded people who acquire, reinforce and shape their preferred narrative~\cite{schmidt2017anatomy,delvicario2017mapping}. In this scenario, dissenting information usually gets ignored~\cite{zollo2017debunking}, thus the effectiveness of debunking, fact-checking and other similar solutions turns out to be strongly limited. 

Since 2013 the World Economic Forum (WEF) has been placing the global danger of massive digital misinformation at the core of other technological and geopolitical risks~\cite{howell2013digital}. Hence, a fundamental scientific challenge is how to support citizens in gathering trustworthy information to participate meaningfully in public debates and societal decision making. However, attention should be paid: since the problem is complex, solutions could prove to be wrong and disastrous. For instance, relying on machine learning algorithms alone (and scientists behind) to separate the truth from the false is na\"if and dangerous, and might have severe consequences.
As far as we know, misinformation spreading on social media is directly related to the increasing polarization and segregation of users~\cite{del2016spreading,zollo2017debunking,quattrociocchi2016echo,zollo2018misinformation}. 
This is a dynamical process whose evolution depends on two factors: i) the engagement of users, that is their attitude and willingness to embrace a given cause or opinion, and ii) the ability of pages to spread a message and have an impact on users, in other words, to engage them. Clearly, these two features are deeply entangled. The aim of this paper is to use the link between these two properties, one an attribute of pages, the other an attribute of users, to obtain a quantitative assessment of both. Such an assessment is the output of the \textit{PopRank} algorithm, its input being the bipartite network \cite{straka2018ecology} pages and users interactions. We build this algorithm in analogy with the Fitness and Complexity algorithm \cite{tacchella2012new}, whose aim is to quantify countries' competitiveness and products' sophistication from the bipartite network of exports. Such an approach has been successfully applied to a number of macroeconomic analyses. For instance, the Fitness of countries has been used to predict GDP growth, showing better results than the state-of-the-art methodologies \cite{cristelli2017predictability}, as stated by a recent Bloomberg View editorial \footnote{www.bloomberg.com/view/articles/2017-10-01/a-better-way-to-make-economic-forecasts}. Moreover, it has been shown that a high value of Fitness lowers the economic threshold countries must face during the escape from the poverty trap \cite{pugliese2017complex}. The other output of the algorithm, the Complexity of products, shows a non trivial dynamics and shapes the respective export markets \cite{angelini2017complex,zaccaria2016case}. This methodology has been used also to investigate the economical features and perspectives of single regions or countries \cite{zaccaria2016case,cristelli2015growth,pugliese2015economic,liao2018comparative,chavez2017economic}.\\
The present paper adopts a similar methodology, introducing an algorithmic assessment of the nodes of the bipartite pages-users network by leveraging its structure. Obviously, the quantities of interest and the observed dynamics are different from the original field of application of the Fitness approach, i.e. macroeconomics. This imposes different methodological choices and, in particular, a different mathematical formulation of the problem and a new algorithm that we name PopRank. The output of this algorithm is an assessment of pages' impact and can be used to predict the users' activity on such pages.\\
The rest of the paper is organized as follows. In the Methods section we describe the database we use to build the pages-users network  and to quantify the future activities of users; we then introduce the PopRank algorithm to measure the Impact of pages and the Engagement of users. In the Results section we show the predictive power of our Impact measure and its dependence on the algorithm parameters; we then analyze the possible effects of users' polarization. We conclude with a discussion of the implications of our results and some possible future applications.

\section*{Methods}
\subsection*{Database}

\begin{figure}[h!]
\centering
\includegraphics[width=0.3\linewidth]{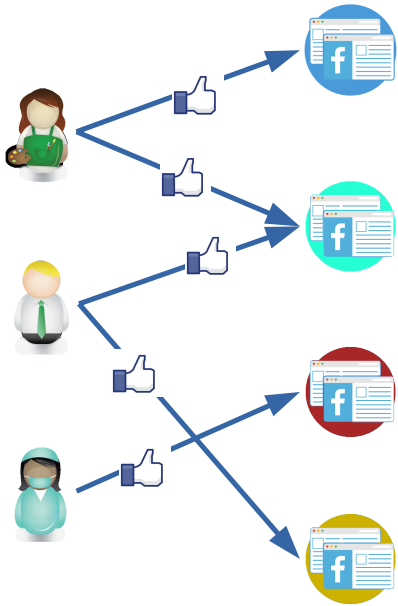}
\caption{Structure of the database used as an input to the algorithms studied in the paper. The database consists in the history of interactions (like, comments) of FB users with FB pages. In this form, it corresponds to a bipartite graph whose edges have a time tag (i.e. when the interaction happened) and therefore can be multiple (each user can comment at different times the same FB page).}
\label{fig:FBdb}
\end{figure}

Our datasets is a subset of the US Facebook dataset of \cite{ZolloPONE2017} analyzed from January $2010$ up to Dicember $2015$ on a per-month basis. \\
The quantities we are interested into are basically three:
\begin{itemize}
\item the monthly Activity \textit{of} a page, that is the number of posts it is producing;
\item the monthly Activity \textit{on} a page, that is the number of comments the page is receiving;
\item the number of users are commenting on a page, possibly divided in groups on the basis of their polarization level (that can be proxied, for instance, by counting how many comments they leave on the same page).
\end{itemize}
All these quantities can be organized in matrices and seen as the weights of three bipartite pages-users networks. The second one - users' activity on a page - will be the input of the PopRank algorithm.
The number of comments per month of the users is a widely distributed quantity, hence we will use, to test the predictive power of the algorithm, the logarithm of the number of their comments \cite{BessiPONE1-2015,BessiPONE2-2015,BessiPOne2016,ZolloPONE2017}. To clean up the noise, we have only considered pages that have been commented at least $5$ times \footnote{Our results are practically unchanged if this threshold is reasonably varies.}; for those pages, we have then considered a sample of $10^6$ users. Notice that both the number of active (i.e. posting) pages and active (i.e. commenting) users can vary month by month.
In total, we have $61$ biadjacency matrices $V^m$, one for each month, where the element $V^i_{u,p}$ is the number of posts received by page $p$ by user $u$ in the $m^{th}$ month. Notice that such matrices can have different dimension; in particular, we observe that both the number of pages and the number of users increase with time. However, since the in the first months of the database the matrices $V^m$ have a very limited number of elements, we consider only the months from the $40^{th}$ onwards. We further divide the remaining months in a \textit{training set} (comprising the months from the $40^{th}$ to the $55^{th}$) and a \textit{test set} (months from the $56^{th}$ to the $61^{st}$). 
\\
In order to build a reliable algorithmic assessment of pages' impact, we want to aggregate such information in one global biadjacency matrix $\bf{V}$, in which we keep only those pages and users which were active in all months.
After this aggregation we come out with a training set composed by a total of $82$ pages and $295$ users. By summing up the monthly matrices, we obtain a global matrix $\bf{V}$ whose elements $V_{u,p}$ indicates the number of comments the users $u$ posted on page $p$ in the time interval from the $40^{th}$ to the $55^{th}$ month; as a further filter, we check that there are no inactive pages (i.e. whose total number of comments in the $15$ month period is less than $5$). 

We now resort to a economic analogy: the matrix $V$ can be considered as representing the amount of time spent by an user on a page; hence, the rows $V_{u,\_}$ indicate how user $u$ distributes his attention on several pages, while the columns $V_{\_,p}$ indicate which are the users "investing" their time on page $p$. Thus, to distinguish where the user concentrate his attention, we compute the binary Revealed Comparative Advantage (RCA) $\bf{M}$ associated with the matrix $\bf{V}$:

\begin{equation}
  M_{up}=
  \left\{\begin{array}{ll}
        1 \qquad\mbox{ if }\quad\dfrac{V_{up}}{\sum_{p'}V_{up'}}\Bigg/\dfrac{\sum_{u'}V_{u'p}}{\sum_{u'p'}V_{u'p'}}\ge1,\\
        0 \qquad\mbox{ otherwise. } 
        \end{array}
  \right.
\end{equation}

Originally introduced in an economical context by Balassa \cite{RCA1965} as the degree of specialization of a country in a product (in that case $V$ contained the total value of the exports of country in a given industrial sector, in a given year), the RCA takes into account possible differences in pages' size and normalizes with respect to such size differences. The binarization procedure discriminates among those pages that are, in this sense, \textit{competitive}: $M_{up}=1$ only if the share of comments of user $u$ on page $p$ with respect to the other pages $u$ comments is greater than the same share of the other users, that is the total share of comments that page receives. Both the RCA calculation and the binarization are standard procedure in the Economic Complexity field: indeed, in addition to the economical or social meaning, they remove high fluctuations from the raw data and greatly improve the signal to noise ratio \cite{hidalgo2007product,hidalgo2009building,CaldarelliPONE2012,tacchella2012new,battiston2014metrics}. In any case, we replicated all the analyses presented in the Results section using both $\bf{M}$ or $\bf{V}$ as input matrices, finding rather similar results, as addressed below.

\subsection*{Algorithm}
As discussed in the Introduction, we would like to extract information from the users' activity on the page adopting a philosophy that is inspired to the approach used by Google's PageRank \cite{Pageetal98}: instead of looking to the fine details of every single page, we build an algorithm to extract the relevant information by exploiting only one carefully chosen variable: while in the case of PageRank variables correspond to the links of a page to other pages, in our case we rely on the users' activity. As a consequence, we will use the bi-adjacency matrix $\bf{M}$ defined in the previous section as the only input of our algorithm. Since the main objective of our approach is to rank pages according to their future impact, or popularity among users, we name our algorithm \textit{PopRank}.\\
Using the same spirit of the Fitness and Complexity algorithm \cite{SciRep}, we aim at building an iterative procedure that assesses at the same time the \textit{Impact} $I_p$ of page $p$ and the \textit{Engagement} $E_u$ of user $u$. To this end, we build a dynamical system $f$
\begin{equation}
\left(\begin{array}{c}
I^{(n+1)}\\
E^{(n+1)}
\end{array}\right)=f\left[\mathbf{M},\left(\begin{array}{c}
I^{(n)}\\
E^{(n)}
\end{array}\right)\right]
\end{equation}
that uses the $\bf{M}$ matrix to evolve some initial conditions $I^{(0)}_p$ and $E^{(0)}$ up to the stationary point $I^{(\infty)}_p$ and $E^{(\infty)}$. The iterative procedure consists in computing $I^{(1)}_p$ using $E^{(0)}$, and then $E^{(1)}_p$ using $I^{(1)}$ and so on, until a convergence criterion is reached. Using extensive numerical simulations, it has been shown that the fixed point of the Fitness and Complexity algorithm is unique \cite{pugliese2016convergence} and independent from the initial conditions; such result holds also for the PopRank algorithm that we introduce in the present paper.\\ 
We now turn our attention both to the explicit mathematical formulation of the PopRank algorithm and to its connection with the users' behavior. A reasonable assumption about the Impact is that pages with higher Impact attract a lot of users, so the total number of users commenting page $p$ should be taken into account. Moreover, we want to weight users according to the inverse of their Engagement, because we want to give importance to those users that are hard to convince. In conclusion, the first equation of our algorithm is
\begin{equation}
I^{(n+1)}_p=\sum_{u}M_{up}\frac{1}{E^{(n)}_{u}}
\end{equation}
where $n$ is the iteration number. In order to estimate, in turn, Engagement from Impact, we adopt a slightly different approach. Suppose that we use the same mathematical expression, that is
\begin{equation}
E^{(n+1)}_u=\sum_{p}M_{up}\frac{1}{I^{(n)}_{p}}
\label{eq2}
\end{equation}
in this case, the meaning would be that the user $u$ is engaged if he comments a lot of pages (this is the meaning of summing over $p$), but that the algorithm would weight more those pages that have lower impact. This is in contrast with the known literature about the polarization of users in social networks \cite{DelVicario554}, that shows that a self-reinforcing mechanism is active, in which users are more and more confined in an echo chamber as they continue to post and comment. As a consequence, we give to our second equation one degree of freedom, an exponent $\alpha$ that regulates how the impact of a page influences users' engagement:
\begin{equation}
E^{(n+1)}_u=\sum_{p}M_{up}\bigg(\frac{1}{I^{(n)}_{p}}\bigg)^\alpha
\end{equation}
For $\alpha=1$ we recover eq.$\,\,$(\ref{eq2}), that would correspond to a simple reformulation of the Fitness and Complexity algorithm.
For $\alpha=0$, the Engagement of the users is not dependent on the impact of the pages, but it is simply given by how many pages are commented, which is a reasonable first-order approximation. We anticipate that we find a better predicting performance for $\alpha<0$. This result indicates that a user's engagement is linked to how many polarizing pages he comments. A negative value in the exponent $\alpha$ agrees with the known literature in misinformation spreading, where it is empirically found that a self-reinforcing process at work\cite{DelVicario554}.\\
Finally, as in \cite{SciRep}, at each iteration we normalize both Impact and Engagement with respect to the respective averages, that we indicate using the symbols $<\dots>$. The algorithm we propose is therefore

\begin{minipage}{0.4\textwidth}
\begin{equation}
\label{eq:Impact}
 \tilde{I}_p^{(n+1)}=\sum_{u}M_{up}\frac{1}{E^{(n)}_{u}}
\end{equation}
\begin{equation}
\label{eq:Superficiality}
\tilde{E}_u^{(n+1)}=\sum_{p}M_{up}\bigg(\frac{1}{I^{(n)}_{p}}\bigg)^\alpha
\end{equation}
\end{minipage}
\hspace{0.2\textwidth}
\begin{minipage}{0.4\textwidth}
\begin{equation}
\label{eq:ImpNorm}
 I_p^{(n)}=\frac{ \tilde{I}_p^{(n)}}{ <\tilde{I}_p^{(n)}>_p}
\end{equation}
\begin{equation}
\label{eq:SupNorm}
 E_u^{(n)}=\frac{ \tilde{E}_u^{(n)}}{ <\tilde{E}_u^{(n)}>_u}
\end{equation}
\end{minipage}
where $I_p$ is the Impact of page $p$, $E_u$ is the Engagement of user $u$. \\The algorithm is iterated until convergence in ranking, using the methodology introduced in \cite{pugliese2016convergence}: at each iteration $n$ we estimate the relative growth rates of the Impacts, and the number of iterations $T(n)$ that one should wait for at least one change in rankings to occur. Our stopping criterion is $T(n)>10^6$, that means that the next change in rankings is expected to happen not before $10^6$ iterations. We point out that this kind of stopping criterion is necessary for this kind of algorithms, in particular for sparse matrices. In fact, depending on the specific structure of the input matrix $\bf{M}$, some (even all but one, in some cases) of the outputs $I_p$ and $E_u$ may converge to 0\cite{pugliese2016convergence}. In such a situation a standard stopping criterion such as $| I_p^{(n)}- I_p^{(n+1)}|<\epsilon$ in not appropriate.
\section*{Results}
\subsection*{Impact predicts users' and pages' activity}
\begin{figure}[h!]
\centering
\includegraphics[width=0.9\linewidth]{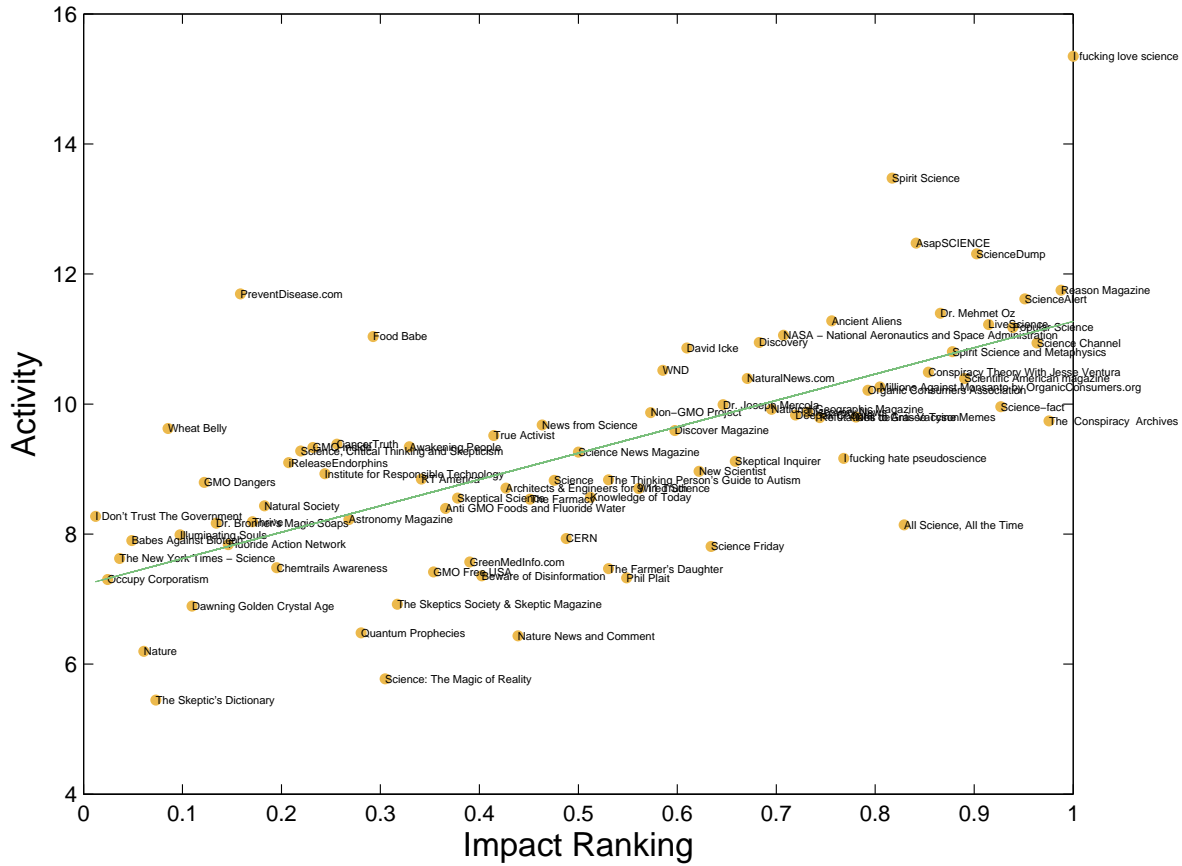}\\
\caption{Future activity (i.e., number of posts) of Facebook pages as a function of Impact ranking. The PopRank algorithm can also predict how many comments will be posted on that page and the number of users will comment its posts. In particular, we show the results for $\alpha=-1/2$.}
\label{fig:scatter}
\end{figure}

We now compare the output of our algorithm, and in particular the Impact $I_p$ of the FB page $p$, with a measure of its \textit{future} performance and activity. In order to test the predictive power of our measure of Impact, we split our dataset, which comprises 22 months of data, in a training set and a test set. While the Impact is computed using the first 16 months of our database, the Activity (defined in the Methods section) is computed using the last 6 months, in such a way that no overlap is present between the training and the test sets. In the following, we will indicate by ranking both the relative ordering of a set of elements and a numerical value corresponding to the normalized position in the ranking of an element: as an example, in the case of $N$ elements, we assign the value $1$ to the element with the highest value (i.e. the first in ranking), while we assign the value $1/N$ to the last element.  In Fig.\ref{fig:scatter} we plot the future Activity of the pages as a function of their Impact ranking for various values of $\alpha$. In particular, we show that the algorithm has better performances for $\alpha=-1/2$ respect to the original algorithm of Fitness and Complexity \cite{SciRep} where  $\alpha=1$.\\ 
As seen from the pictures, there is a high correlation among the two quantities $A$ and $I$; such a correlation, measured as the explained variance $R^2$ of a linear regression model, reaches a maximum value $R^2 \sim 0.46$ for $\alpha \approx -1/2$. Using a t-test, one can show that this correlation is statistically significant (p-value $\approx 10^{-12}$). We use the Impact ranking, rescaled in such a way that the page with the highest Impact has and Impact ranking equal to 1. The use of the ranking will be fundamental to compare the results of the algorithm as a function of the exponent $\alpha$. In fact, for some values of $\alpha$ some numerical values of both Impact and Engagement converge to zero, as expected for some typologies of sparse input matrices $\bf{M}$ \cite{pugliese2016convergence}. In these cases, as already seen in \cite{pugliese2016convergence}, only the ranking and not the values of Impact and Engagement is meaningful. In order to be able to compare the outputs of different versions of the algorithm (i.e., different exponents) we decided to use the rankings for all these versions. We point out that the results shown in Fig.\ref{fig:scatter}, and in particular the high correlation between the Impact and the future Activity, holds also when the actual values of the Impact and not the rankings are taken into account. We repeat this analysis to predict also the activity \textit{on} the page, that is the number of comments leaved and the number of users that are commenting the posts of that page. We find similar results for predicting activity \textit{on} and \textit{of} the page also when users are separated according to their polarization level (see next section).\\

\begin{figure}[h!]
\centering
\includegraphics[width=0.9\linewidth]{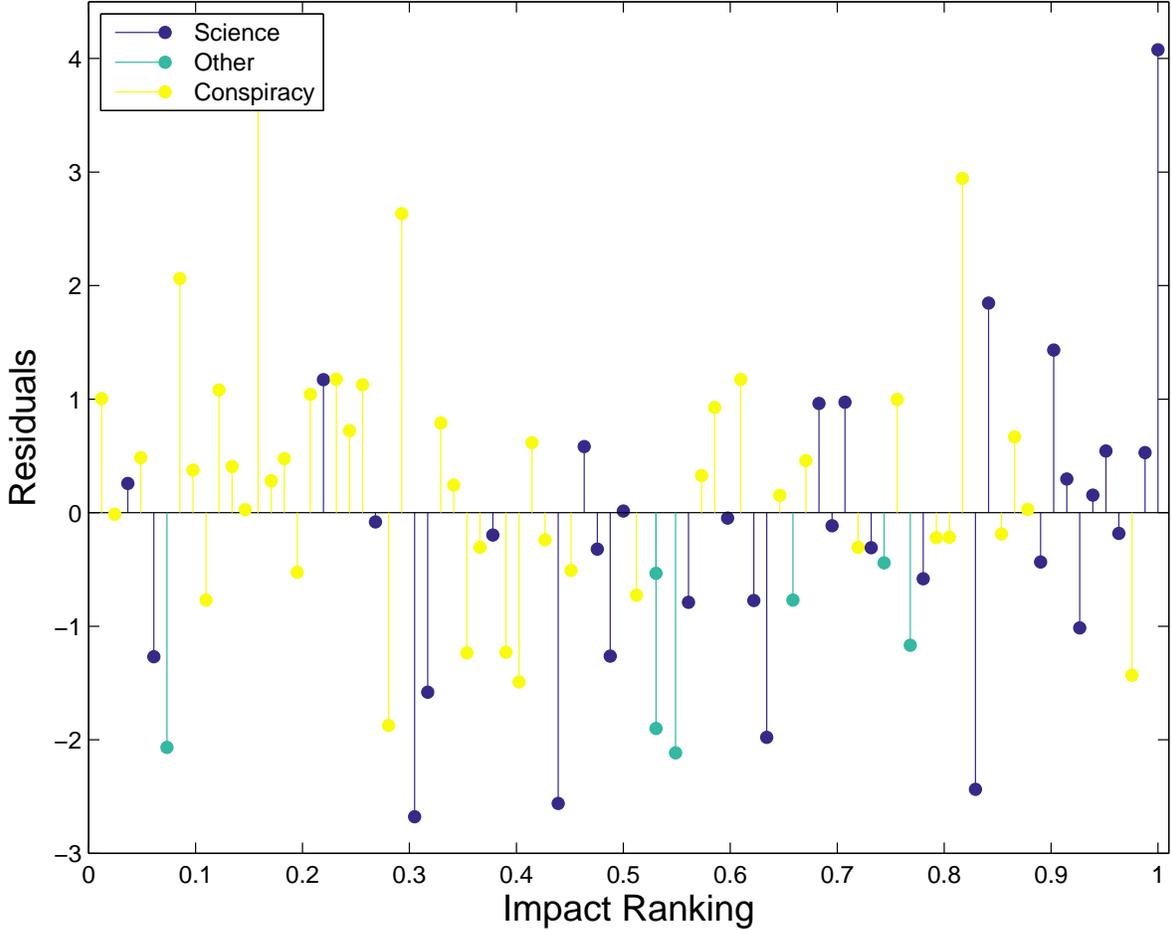}
\caption{Residuals of the linear fit in Fig.\ref{fig:scatter} as a function of the Impact Ranking. Scientific and conspiracy pages show a similar behavior with respect to the residuals. On the contrary, the Impact ranking shows a slight discriminative power since, on average, conspiracy pages have a lower ranking respect to scientific ones. }
\label{fig:stem}
\end{figure}

As discussed in the introduction, previous studies have found a substantial symmetry between the polarization dynamics regardless of the specific conveyed information. Our dataset contains both scientific and conspiracy pages, so a natural question is whether the belonging to one of these groups affects the results of our analysis. In order to investigate this possible dependence we compute the residuals of the linear fit shown in Figure \ref{fig:scatter} and we plot them as a function of the Impact ranking. In the resulting plot, shown in Figure \ref{fig:stem}, we use different filling colors on the basis of the group each page belongs to. One can easily see that the residuals do not depend on the group, that is, our algorithm is performing more of less in the same way for both scientific and conspiracy pages. \\
The same plot shows also that the Impact ranking has almost no discriminative power between the two groups: conspiracy pages tend to occupy on average a slightly lower positions in the ranking with respect to scientific pages. We believe this feature should be quantified in some way, we plan to perform this analysis in a future work.\\

\begin{figure}[h!]
\centering
\includegraphics[width=0.9\linewidth]{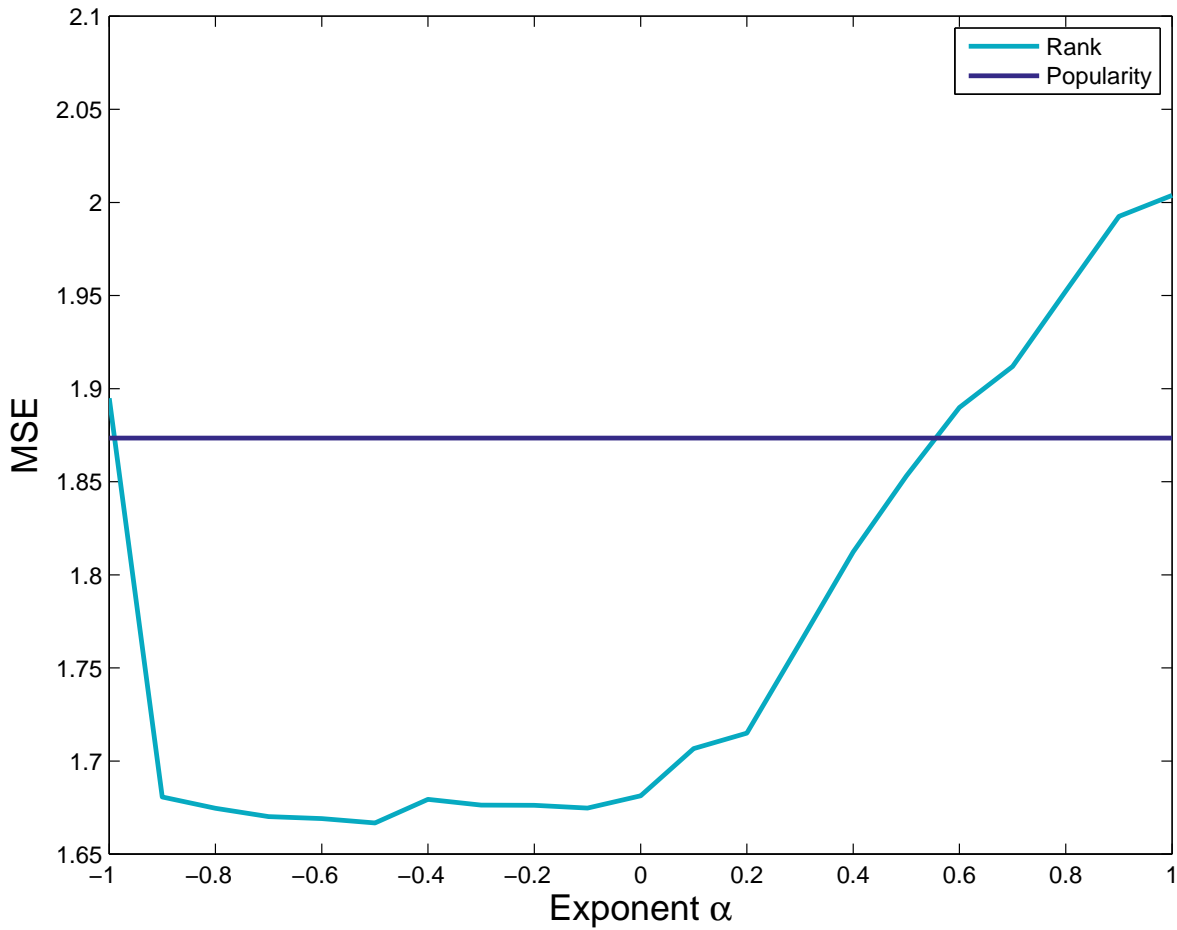}
\caption{Mean squared error in predicting the Activity using the simple Popularity measure or using the Impact, as a function of the exponent $\alpha$ in the PopRank algorithm. Negative exponents give better results.}
\label{fig:mse}
\end{figure}

Now we test how the predictions given by our assessment of pages' Impact depend on the exponent $\alpha$ in Eq.\ref{eq:Superficiality}. We have considered various ways to quantify such predictive power, such as R-squared, p-value associated to the hypothesis of independence, and Mean Squared Error (MSE). All these quantities give the same qualitative results. In the following we will focus on the $MSE$, defined as
\begin{equation}
MSE=\frac{1}{P-2} \mathlarger{\sum}_{p} \big(A_p-\hat{A}_p\big)^2
\label{mse}
\end{equation}
where $P$ is the number of pages, $A_p$ is the Activity of or on page $p$ and $\hat{A}_p$ is its estimation coming from the linear fit. The factor $\frac{1}{P-2}$ takes into account the fact that the effective degrees of freedom are lowered by the presence of the two parameters (intercept and angular coefficient) of the linear model. In Fig.\ref{fig:mse} we show the $MSE$ as a function of the exponent $\alpha$. One can easily see that negative exponents give better predictions, being the associated mean error lower (light blue line). This means that users' Engagement should be computed weighting more those pages that have higher impact. In other words, there is a self-reinforcing process at work, in which those users that are easier to convince interact more with those pages that have a higher impact. This is at odds with respect to Economics: in the Fitness and Complexity algorithm, in fact, when one computes the Complexity of a product more weight is given to those countries that have a lower Fitness \cite{tacchella2012new}, a situation that corresponds to  $\alpha=1$ in our formulation (see Eq. \ref{eq:Superficiality}).\\
For a comparison, we show also the $MSE$ associated with another possible predictor, the \textit{Popularity} of the page computed as the sum $\sum_u M_{up}$ (dark blue line, obviously independent from the value of the exponent). Notice that the Fitness and Complexity algorithm (which is recovered in the case $\alpha=1$) not only underperforms with respect to $\alpha<1$ exponents, but also with respect to the simpler measure of popularity. This fact stresses the intrinsic difference between the two applications, development economics on one hand, social science and information spreading dynamics on the other hand. \\

\subsection*{Polarization analysis}

\begin{figure}[h!]
\centering
\includegraphics[width=0.9\linewidth]{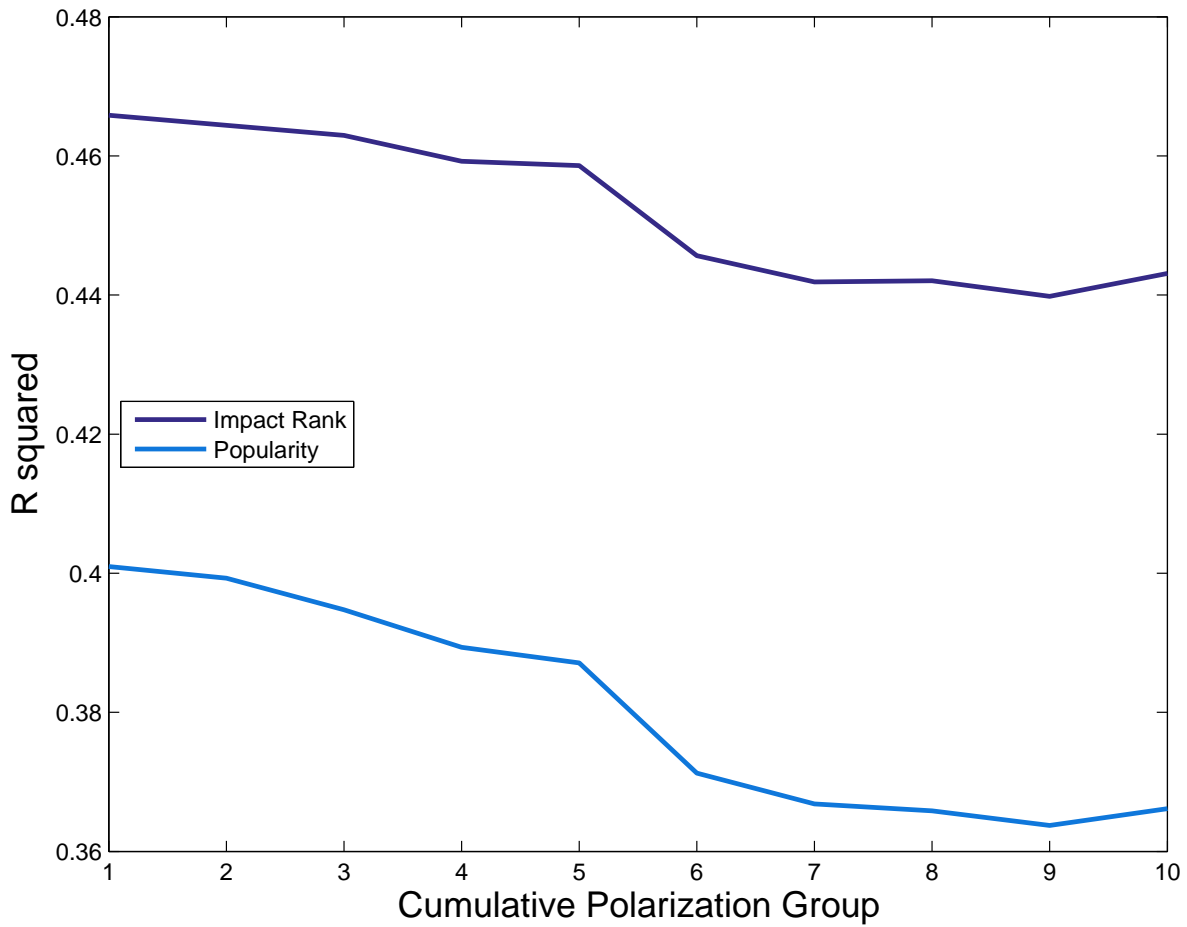}
\caption{The correlation between Impact and future Activity is roughly independent from users' polarization level. We divide users according to their polarization and we count the number of users, belonging to a given group, that comments a given page. The PopRank algorithm can predict such values with similar performances across the groups, and always overperforming a simpler measure of Popularity.}
\label{fig:perc}
\end{figure}

We now analyze the possible dependence of the algorithm performance on users' polarization. In order to quantify how much a page engages polarized users, following the results of \cite{BessiWWW2015}, we count how many comments each user posts on a given page and we divide this number by her total number of comments. This ratio $x$ is a proxy of users' polarization  \cite{BessiWWW2015}. We then consider, for each page, $10$ different groups of users on the basis of their polarization ratio: greater or equal to $x^\star=0.1, 0.2 \dots 1$, where $x=1$ means totally polarized users, that is to say users that comment only that page. The number of comments coming from each one of these cumulative polarization groups depends on $x$ and is a proxy of the page's impact on users showing different degrees of polarization. In practice, we repeat our analysis for each one of these groups using as the Activity on a given page the number of users commenting that page. As we show in \ref{fig:perc}, the prediction performance of PopRank is substantially independent from the polarization group. If any, it performs better on lower polarization levels, that is, taking into account not only the polarized users but also the ones that comment different pages.  

\section*{Discussion}

In this paper we have introduced a novel algorithm, called PopRank, to rank both Facebook pages and users on the basis of their mutual interaction. To do so we have built a bipartite network whose links indicate that a given user is commenting the posts of a given page more than a suitable average. The bi-adjacency matrix of the network is the only input of PopRank, whose output is a quantitative assessment of pages' Impact and users' Engagement. In particular, we compute the two quantities one as a function of the other, iterating a system of coupled equations up to convergence. The general idea is that pages with a strong Impact are commented by many users with a low Engagement, and users have a high Engagement if they comment many pages with a high Impact. The Impact can be used to successfully predict the activity of and on users on a given page with a six months time delay.
This result is robust with respect to reasonable variations of the algorithm's only parameter $\alpha$; in particular, the effectiveness of negative values of $\alpha$ indicates that more engaged users act on higher impact pages. 
Moreover, we find that high Impact pages engage users regardless of their polarization.
These results have been obtained by analyzing Facebook pages without any discrimination based on their informational content. This means that, for instance, scientific dissemination and fake news are processed in the same way and show the very same behavior: in particular, the relationship between their Impact and the future activity of their users is practically the same. This finding confirms the substantial symmetry between pages (and users) of different opinion, regardless of the possible veracity, if any, of the conveyed information.

\section*{Acknowledgements}

AZ, AS, WQ and LP acknowledge the Italian PNR project CRISIS-Lab; AS and WQ acknowledge the CNR project AMOFI.
The funders had no role in study design, data collection and analysis, decision to publish, or preparation of the manuscript.
The authors would like to thank Giulio Cimini, Emanuele Pugliese and Andrea Tacchella for early discussions.

\section*{Author contributions statement}

All authors designed the study; A.Z. performed the analysis; A.Z. and M.D.V. contributed data and methods; all authors analyzed and interpreted the results; all authors wrote and revised the paper. 

\section*{Additional information}

\textbf{Competing financial interests} The authors declare no competing financial interests.

\end{document}